\documentclass[journal=jacsat,manuscript=article]{achemso}

\usepackage{mciteplus}
\usepackage[version=3]{mhchem} % Formula subscripts using \ce{}

\author{Wayne Yang}
\email{wayne.yang2@mail.mcgill.ca}
\altaffiliation{These authors contributed equally}
\author{Yuning Zhang}
\email{yuning.zhang2@mail.mcgill.ca}
\altaffiliation{These authors contributed equally}
\author{Michael Hilke}
\author{Walter Reisner}
\phone{+123 (0)123 4445556}
\fax{+123 (0)123 4445557}
\affiliation[McGill University]
{Department of Physics and RQMP, McGill University, Montreal}

\title[An \textsf{achemso} demo]
  {Dynamic Imaging of Au-nanoparticles via Scanning Electron Microscopy in a Graphene Wet Cell}

\abbreviations{IR,NMR,UV}

\begin{document}
%%%%%%%%%%%%%%%%%%%%%%%%%%%%%%%%%%%%%%%%%%%%%%%%%%%%%%%%%%%%%%%%%%%%%
%% The manuscript does not need to include \maketitle, which is
%% executed automatically.  The document should begin with an
%% abstract, if appropriate.  If one is given and should not be, the
%% contents will be gobbled.
%%%%%%%%%%%%%%%%%%%%%%%%%%%%%%%%%%%%%%%%%%%%%%%%%%%%%%%%%%%%%%%%%%%%%
\begin{abstract}
High resolution nanoscale imaging in liquid environments is crucial for studying molecular interactions in biological and chemical systems.  In particular, electron microscopy is the gold-standard tool for nanoscale imaging, but its high-vacuum requirements make application to in-liquid samples extremely challenging.   Here we present a new graphene based wet cell device where high resolution SEM (scanning electron microscope) and Energy Dispersive X-rays (EDX) analysis can be performed directly inside a liquid environment.   Graphene is an ideal membrane material as its high transparancy, conductivity and mechanical strength can support the high vacuum and grounding requirements of a SEM while enabling maximal resolution and signal.  In particular, we obtain high resolution ($<$ 5 nm) SEM video images of nanoparticles undergoing brownian motion inside the graphene wet cell and EDX analysis of nanoparticle composition in the liquid enviornment.  Our obtained resolution surpasses current conventional silicon nitride devices imaged in both SEM and TEM under much higher electron doses.  
\end{abstract}

\section{Introduction}

Imaging in a liquid environment is important across a wide range of research fields from physics to biology. The nanoscale imaging of such systems drives new insights in molecular and biological theory \cite{ackerley2006experiences,ishii2005large}. New experiments enabled by wet-cell technology include live imaging of antibodies and bacteria to understand immune response, and in-situ imaging of crystals to understand growth kinetics \cite{niu2013seed}. Most of the imaging is performed with electron microscopy such as SEM (Scanning Electron Microscope) or TEM (Transmission Electron Microscope)\cite{kolmakov2011graphene,tsuda2011sem}. In particular, SEMs are widely available and accessible to most researchers. While SEMs offer quick and high resolution nanoscale (2-10nm) imaging, the high vacuum operation conditions (\textless $10^{-4}$ Torr) of these instruments make the imaging of liquid environments challenging \cite{bogner2007history}. Systems that operate at high pressures such as environmental SEMs (E-SEMs) are specialised tools requiring the use of water vapour to purge and replace air in the specimen chamber.  Moreover, the electron beam in such systems scatters from the introduced vapour resulting in limited resolution \cite{muscariello2005critical,peckys2013detection,donald2003use}.  

Conventional wet cells are based on sealing liquid samples behind a 30-150\,nm silicon nitride window\cite{yang2014situ,mueller2013nanofluidic}. While this approach has proved effective, the resolution is fundamentally limited by the necessity of using relatively thick nitride membranes.  Experiments have obtained a resolution of only around 20 nm for a membrane thickness of 50 nm in an SEM \cite{thiberge2004scanning}. The fabrication of thinner nitride windows with thickness below 50 nm is challenging, requiring special techniques to control the etching rate and achieve etching uniformity \cite{pr2011}.  As the nitride membrane becomes thinner, the windows become too fragile to handle. Silicon nitride wafers are also electrically insulating, requiring the sputtering of a thin layer of conductive material such as gold for electrical leads or to ground the sample\cite{firnkes2010electrically}.  Ultimately, nitride based windows cannot be extended to thicknesses below a few nanometres. This is a very crucial technical limitation, limiting not just resolution but signal.  For example, the need for relatively thick nitride windows obviates application of standard SEM techniques such as Energy Dispersive X-Ray (EDX) due to the absorption of signal by the thick membrane.

Here we present a graphene wet cell for SEM imaging under a high vacuum environment. Graphene is an atomically thick layer of carbon atoms (0.34\,nm thickness)~\cite{cooper2012experimental} with exceptional properties including high mechanical strength, high thermal and high electrical conductivity.   Graphene's atomic thickness makes the material an optimal imaging window enabling maximum resolution and signal. In particular, graphene allows for the collection of low energy secondary electrons as opposed to just backscattered electrons  performed in most SiN wet cell imaging studies \cite{nishiyama2010atmospheric}. This greatly improves the signal and resolution of the images.” Graphene's mechanical strength prevents breakage of $\sim$5\,$\mu$m membranes under vaccum conditions.  Graphene's high thermal conductivity allows excess heat generated from the beam to dissipate quickly without damaging the sample.  Finally, graphene's high electrical conductivity obviates the need for an additional metal coating for grounding. The graphene membrane also provides convenient electrical leads for voltage and current inputs for adding electrical bias in experiments.  Previous groups have used graphene oxide membranes for imaging~\cite{krueger2011drop}. However it is challenging to control the homogeneity in the graphene oxide membrane across the window and, at around 20\,nm thick, they are comparable in thickness to nitride.   Using chemical vapour deposition (CVD) with carefully controlled growth conditions we can ensure that there is a single layer graphene membrane \cite{li2009large,massicotte2013quantum}. 
 
Our single-layer graphene wet cell device enables dynamic imaging in a SEM.  In particular, we observe Brownian dynamics of Au-NP's transiently binding and unbinding at the surface of the graphene.  While Brownian motion of Au-NP's has been observed previously in a TEM using a graphene sandwich assay,  developing a molecular in-liqiuid imaging capability in an SEM has key practical and fundamental benefits\cite{chen20133d}.  SEM's are more available, cheaper and more versatile tools that permit introduction of much larger samples. For example, large (1-10 cm size) micro/nano fluidic devices could be easily introduced into an SEM and wet cell imaging could then be performed as part of routine device operation.  In particular, as there is no constraint on sample thickness in an SEM, an SEM-based wet cell can incorporate much deeper fluidic channels without loss of signal, significantly simplifying wet-cell design. Moreover, additional sample material can be potentially pulled in from deeper in the cell.  For example, we show that continuous scanning attracts Au-NP's to the graphene interface.  Finally, SEMs are outfitted with a wide range of surface characterisation tools (for example, EDX).  We show that, using our graphene wet cell device, these tools can then be adapted to study the wet cell environment.  As an example, we able to obtain an EDX spectrum of Au NPs in liquid.

\section{Sample preparation}

\begin{figure}[H]
\centerline{\hbox{ \hspace{0in}}
\resizebox{1\columnwidth}{!}{\includegraphics{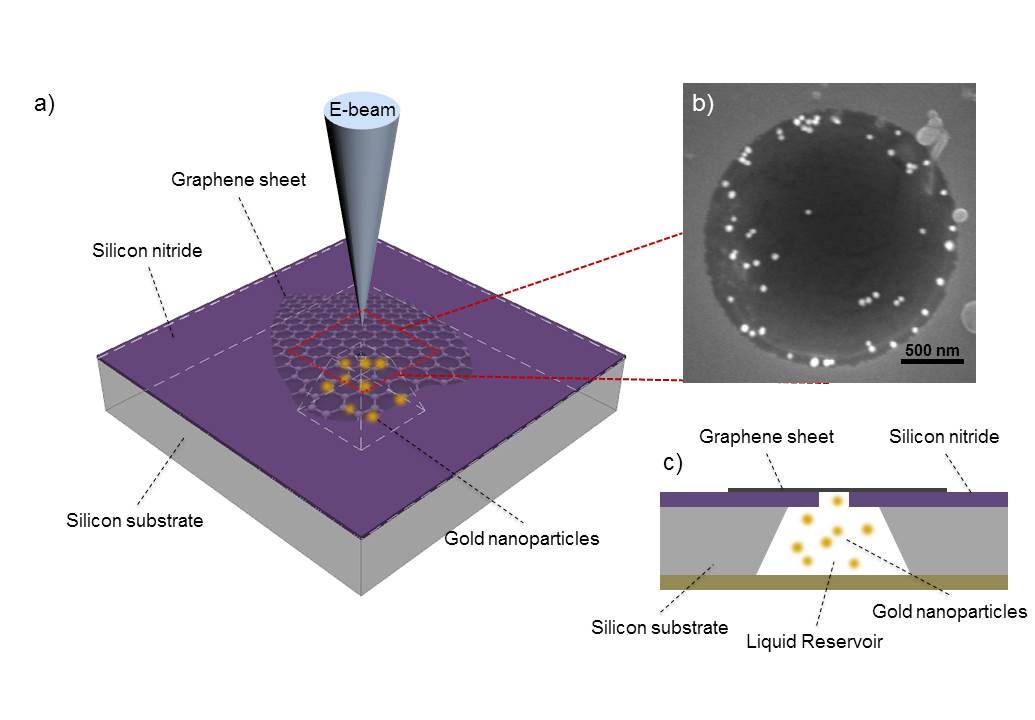}}}
    \caption{\small{a) A schematic of our graphene wet cell device. b) SEM image of the liquid environment imaged
through the graphene membrane micropore. The graphene is positioned on top of a circular aperture
etched through the SiN membrane. c) Schematic of device as viewed from the side. The liquid sample, held in the 400\,$\mu$m fluid reservoir sandwiched between the graphene membrane and kapton tape,
consists of deionized water with Au nano particles.  The figure is not drawn to scale.} }
\label{graphenewetdevice}
\end{figure}

Our fabrication process is divided into three steps, the fabrication of the silicon nitride substrate, the growth and transfer of the graphene and the wetting and sealing of the device for SEM imaging.  An illustration of the device is shown in Figure ~\ref{graphenewetdevice}.  

The first step is substrate fabrication. Our substrate is a 400$\mu m$ thick (110) silicon wafer coated with a 180\,nm thick nitride membrane and divided into 2$\times$2\,mm dies. The wafer was patterned with photolithography and etched in KOH from the back to produce a 70$\times$70\,$\mu$m residual nitride membrane in the middle of each die.  The KOH etched apertaure also serves as a reservoir for the liquid sample.  Lastly, a 2\,$\mu$m diameter hole was etched through the middle of the free standing nitride membrane to form the graphene viewing window.

Graphene was grown using Chemical Vapour Deposition (CVD) on a 25 $\mu m$ thick copper foil with a growth temperature of 1050 $^{o}$ C at a pressure of 100\,mTorr and a flow of 4 sccm of CH$_{4}$\cite{yu2011raman}. Our custom-built CVD system is based on a vertical furnace. Two gas tubes feeds into the top of a 2.5\,cm wide vertical quartz tube to provide the flow of gases. The quartz tube is lowered into the oven during the growth and the growth time is approximately 1 hour. Before the growth, the copper foil was first annealed in a flow of 12 sccm of hydrogen for an hour to strip the oxide layer on the foil. The CVD synthesized graphene was then spin coated with a thin supporting layer of polymethylmethacrylate (PMMA) layer and the Cu substrate was etched away in a solution of 0.1\,M ammonium persulfate ((NH$_{4}$)$_{2}$S$_{2}$O$_{8}$).  The sample was transferred by inserting a glass slide into the ammonium persulfate solution, using the slide to scoop out out the freely floating graphene membrane and depositing the graphene bearing slide into a beaker of de-ionised water. To completely remove the ammonium persulfate residues, the sample was transferred into another clean beaker of de-ionised water before being transferred onto the top side of the silicon nitride wafer sample to cover the 2\,$\mu$m holes. Graphene produced using the same growth conditions was transferred onto Si0$_2$ wafers for Raman spectroscopy to confirm that the graphene was indeed monolayer. 

Finally, the sample was ready to be wetted and sealed. Gold nanoparticles (Au-NP's) 20 and 50\,nm in diameter were used to characterise the fluid cell.  We chose Au particles as they are commercially available in a wide variety of sizes and can potentially be used as conductive biological labels \cite{jain2007nanoparticles}.  The Au-NP's were diluted 1:20 from stock solution in DI and then the nanoparticle containing solution was degassed for an hour.  Degassing was crucial to ensure proper wetting and to decrease the formation of gas bubbles during imaging.  After degassing, several microliters of solution was pipetted into the reservoirs and the wafer sample was sealed with Kapton tape on the back side. The device was then rinsed in acetone and isopropanol to dissolve the PMMA supporting layer on the graphene. The imaging of the device was then done using a FEI-F-50 SEM in the standard high vacuum mode at 10$^{-6}$ Torr using a secondary electron detector of the Everhart-Thornley type. The graphene membrane remained intact at this operating pressure of 2.2 $\times$ 10$^{-6}$ Torr. The primary electron energy used for imaging is 10 KeV.  Under these imaging conditions, the escape depth of secondary electrons in water should be in the order of 10s of nm\cite{winter2006photoemission}.
 
\section{Observation of Nanoparticle Dynamics}

\begin{figure}[h!]
\centering
\includegraphics[scale=0.5]{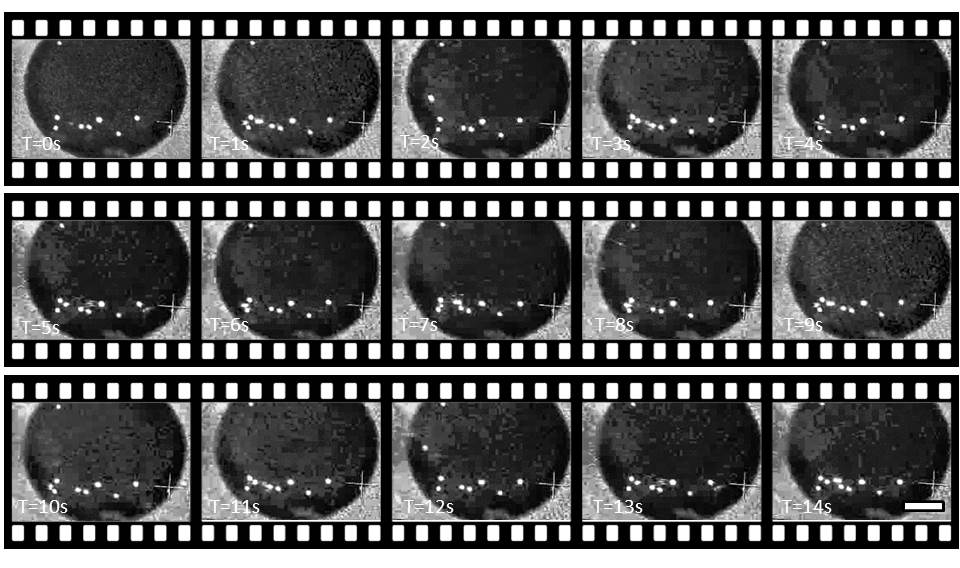}
    \caption{\small{Image time-series showing Au-NP dynamics in our graphene wet-cell device. Beads are observed to be diffusing in and out of contact of the graphene nanopore. The white bar indicates 500 nm.} }

\label{combinefig}
\end{figure}

   While many Au-NP’s are non-specifically bound to the membrane, we observe Brownian dynamics of Au-NP’s floating in solution below and undergoing transient interactions with the membrane.  These dynamics are recorded over several minutes using a screen capture program. Figure \ref{combinefig} gives an example of bead motion.  Beads are observed to be diffusing in and out of
contact with the membrane surface, confirming that they are indeed contained in a liquid environment. The particle trajectories are recorded using a custom tracking program \cite{mueller2013nanofluidic}.

\begin{figure}[H]
\centerline{\hbox{ \hspace{0in}}
\resizebox{1\columnwidth}{!}{\includegraphics{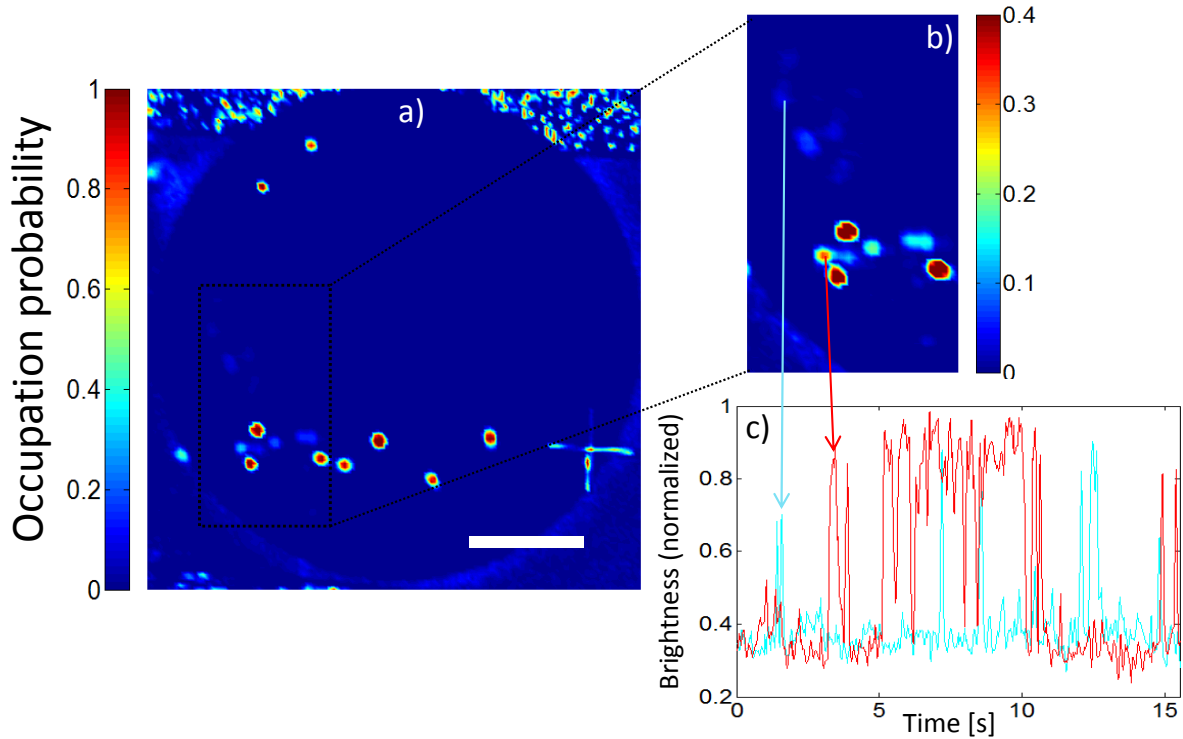}}}
    \caption{a) Plot of the integrated (over 15s) normalized nanoparticle occupation probability across the graphene membrane. The nanoparticle occupation is defined as a brightness of 70 \% or more. Some spots show an occupation of unity, meaning that beads are bound to the membrane at these positions for the entire duration of the movie. b) Zoomed image of the upper left corner with arrows indicating the positions of the time traces in (c). Each pixel shown corresponds to an integrated area of 30$\times$30nm$^2$ at a frame rate of 29Hz. The scale bar denotes 500 nm in length. }
\label{occupation}
\end{figure}

    In the absence of confinement, the gold nanoparticles are expected to undergo Brownian motion in water, characterized by a diffusion constant :
    \begin{equation}    
  D=\frac{k_{b}T}{6\pi\eta r} 
  \end{equation}
  
where k$_{b}$ is the Boltzmann 's constant, T is the temperature (300 K), $\eta$ is the viscosity of DI water (1$\times$10$^{-3}$ Pa$\cdot$ S), r is the radius of the beads (25 nm). For our image frame rate of 1/t$_s$=29Hz, this leads to a corresponding mean diffusion length of $L_D=\sqrt{Dt_s}\simeq 550$\,nm at room temperature (300 K). Hence, within one image frame the particles are expected to approximately hop 1/4 of the length of the nanopore. Figure \ref{combinefig} suggests that we indeed see fluctuations on that scale. However we also observe two additional types of behaviours. Particles can be permanently bound to the membrane over the course of the imaging time and can also diffuse in and out of contact with the membrane, interacting transiently with what appears to be ``sticky sites". This sticking behaviour can be quantified by a plot of occupation probability $p(x,y)$. The occupation probability is taken by integrating the total number of frames a bead appears at a certain location normalised over the total number of frames of the video. Figure \ref{occupation} shows the occupation probability for the same device with a spatial resolution of 10 nm and time resolution of 25 ms, clearly indicating the existence of strong trapping sites that permanently bind beads and weaker trapping sites that give rise to transient interactions.  

\begin{figure}[h]
\centerline{\hbox{ \hspace{0in}}
\resizebox{1\columnwidth}{!}{\includegraphics{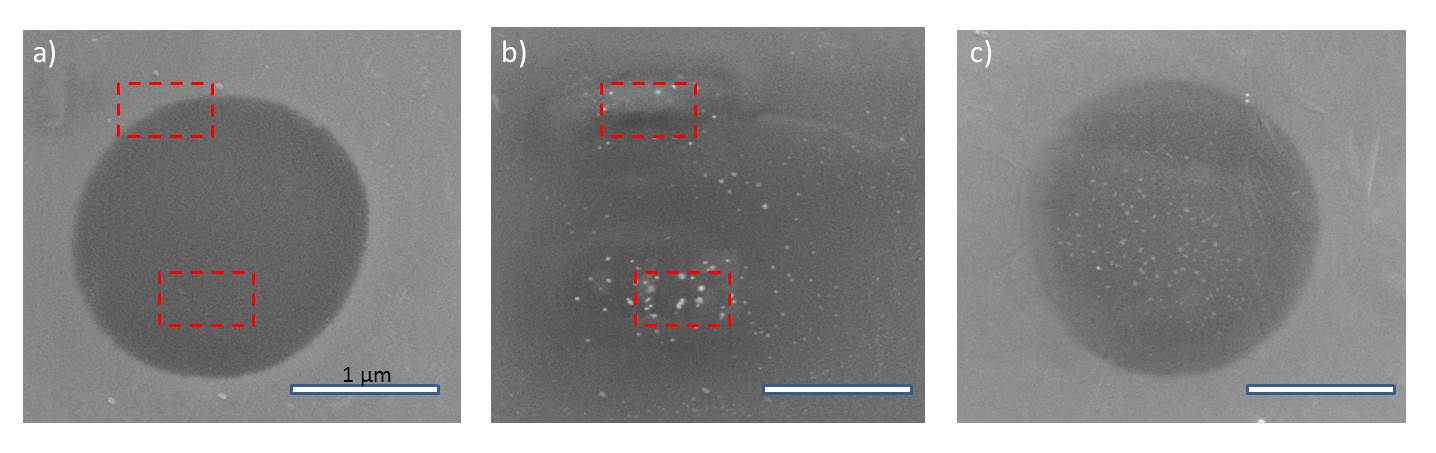}}}
    \caption{\small{a) The graphene pore before imaging in SE mode. The red boxes indicate the areas selected to be scanned for 2 minutes. b) The image of the pore after scanning. c) Image of a graphene membrane on a different device that has been repeatedly scanned by the electron beam. Note that the beads are drawn to the surface. The white scale bar indicates 1 $\mu$m. } }
\label{stickybeads}
\end{figure}

  The non-uniformity of the occupation probability suggests that the graphene membrane varies with regards to its physical and chemical reactivity towards nanoparticles.  One possible source of non-unformity are the existence of grain-boundaries in the graphene layer \cite{huang2011grains}.  These grain boundaries are imperfections in the graphene lattice due to differently orientated growth directions. The grain boundaries from previous studies are spaced roughly the same distance apart ($\sim 1 \mu$m) as the observed sticky sites.  Another possible source of non-uniformity is the presence of graphene ``wrinkles" arising from the growth conditions on the inhomogeneous surface of the copper foils \cite{li2009large}. The wrinkles form valleys in the graphene sheet allowing beads to be drawn in through attraction by van der Waals forces (which has also been observed in other wet cell applications)\cite{moghimi2003stealth}. To reduce this effect, we repeated the experiment with PEG (polyethylene glycol) coated Au beads as shown in Figure \ref{stickybeads}.  Indeed we observed a suppression of the adhesion of Au particles to the graphene membrane with a reduction of the density of stuck beads upon imaging ~\cite{upadhyayula2012coatings}. 

Our dynamic SEM imaging capability, performed over a deep sample reservoir, enables us to demonstrate that continuous scanning draws beads to a scanned region of the membrane. We selected two areas (0.7$\times$0.2$\mu$m) on the graphene membrane, marked by the dashed red boxes in Figure \ref{stickybeads} a).  The area was then scanned continuously at 5\,KeV for 2 minutes.  Beads were observed to diffuse onto the graphene membrane in those areas (Figure \ref{stickybeads} b). Note that Figure \ref{stickybeads} b) is exactly at the same spot as Figure\ref{stickybeads} a). The image looks different because the graphene membrane was observed to be deforming from the continuous scans.  In Figure \ref{stickybeads} c), we scanned the entire pore on a different device for several minutes to draw beads onto the graphene membrane.  This dynamic beam-induced attraction of the beads might arise from electrostatic charging of the membrane, possibly related to chemical modification of local impurities such as PMMA residues (resulting in charge trapping)\cite{lin2011graphene}.  In addition, the PMMA from the supporting layer in the transfer process may not be totally removed during acetone-based dissolution process. Finally, space charge transiently deposited in a nanoscale region beneath the graphene by either electron depositon or secondary electron generation might induce polarization forces on the beads \cite{kochat2011high}. The stability of the graphene membranes greatly varies due to these effects. We observed cells that were stable for 2-15 minutes under 10 KeV. Future cells can be improved by optimizing the imaging conditions and reducing contamination of the graphene membrane.

\section{Energy Dispersive X-ray Spectroscopy}

The use of EDX in a wet cell could potentially allow for positive chemical identification of elements in a liquid environment.    EDX, however, cannot be performed in a standard silicon nitride wet cells due to the thickness of the nitride layer that absorbs emitted radiation.  Here we show that graphene membranes enable EDX-based analysis in liquid enviornments.  

\begin{figure}[H]
\centerline{\hbox{ \hspace{0in}}
\resizebox{1\columnwidth}{!}{\includegraphics{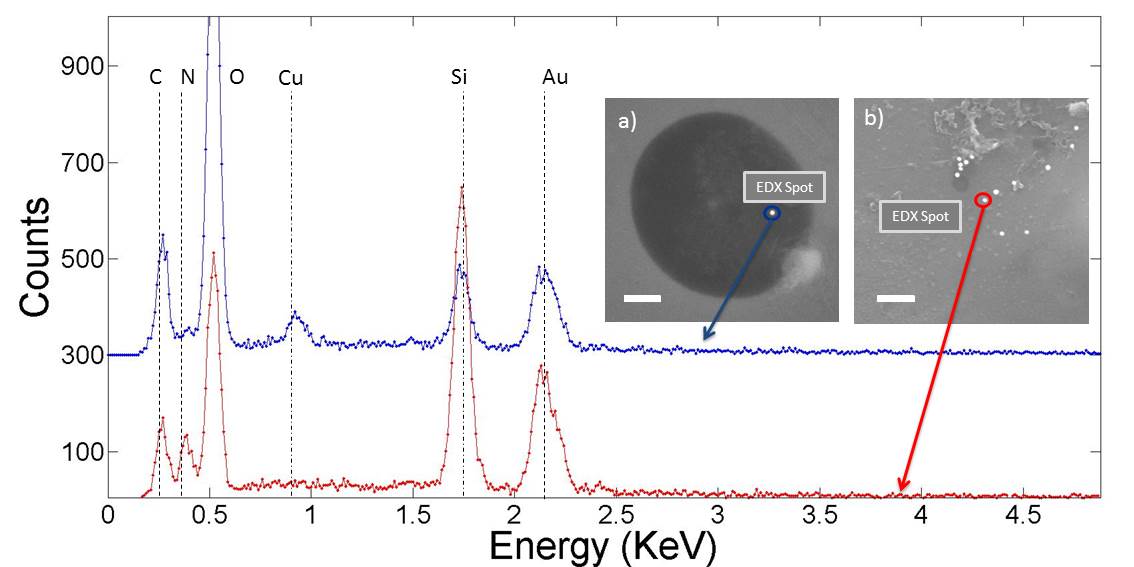}}}
    \caption{EDX spectra for a bead underneath the membrane (blue line) and at a control location away form the membrane and on top of the nitride film (red line).  Both spectra were integrated over 30\,s and the bead-in-liquid spectrum is offset by 300 counts for ease of comparison.   Inset a)  Secondary electron image of location of bead-in-liquid.  Inset b) Secondary electron image of control bead. Both scale bars indicate 500\,nm.} 
\label{EDX}
\end{figure}

We attracted diffusing Au NP's with the electron beam to the surface of the graphene membrane and performed EDX.  Figure \ref{EDX} shows spectrums  taken at two different locations.  The first EDX location (Figure~\ref{EDX} a)) is for a graphene bead located underneath the membrane layer.  The second location (Figure~\ref{EDX} b)) is for a Au-NP resting on top of the graphene-silicon nitride wafer away from the membrane.  To ensure that the EDX spectra correspond to beads in liquid and not on the surface of the membrane, we performed EDX only on beads that had freely diffused onto the surface of the graphene membrane during imaging and were stuck there during the EDX.  Remarkably, we only see a 30\% reduction in the integrated intensity for the Au signal under the graphene vis-a-vis the control spectrum. The source of the attenuation may be due to absorption of the signal by surrounding water or contamination deposited by the electron beam during the EDX measurement.  Despite the attenuation, we are still able to positively identify the in-liquid particle composition.  We also observe a much lower but non-zero silicon peak coming from location (a). The peak arises from the silicon background scattering from the hole edges.  In addition, we observe a weak copper peak on the suspended graphene membrane, likely arising from copper used in the growth process that is not completely removed.  We were able to perform multiple EDX measurements without any degradation of the graphene membrane.  

 \section{Resolution}

\begin{figure}[H]
\centering
\includegraphics[scale=0.45]{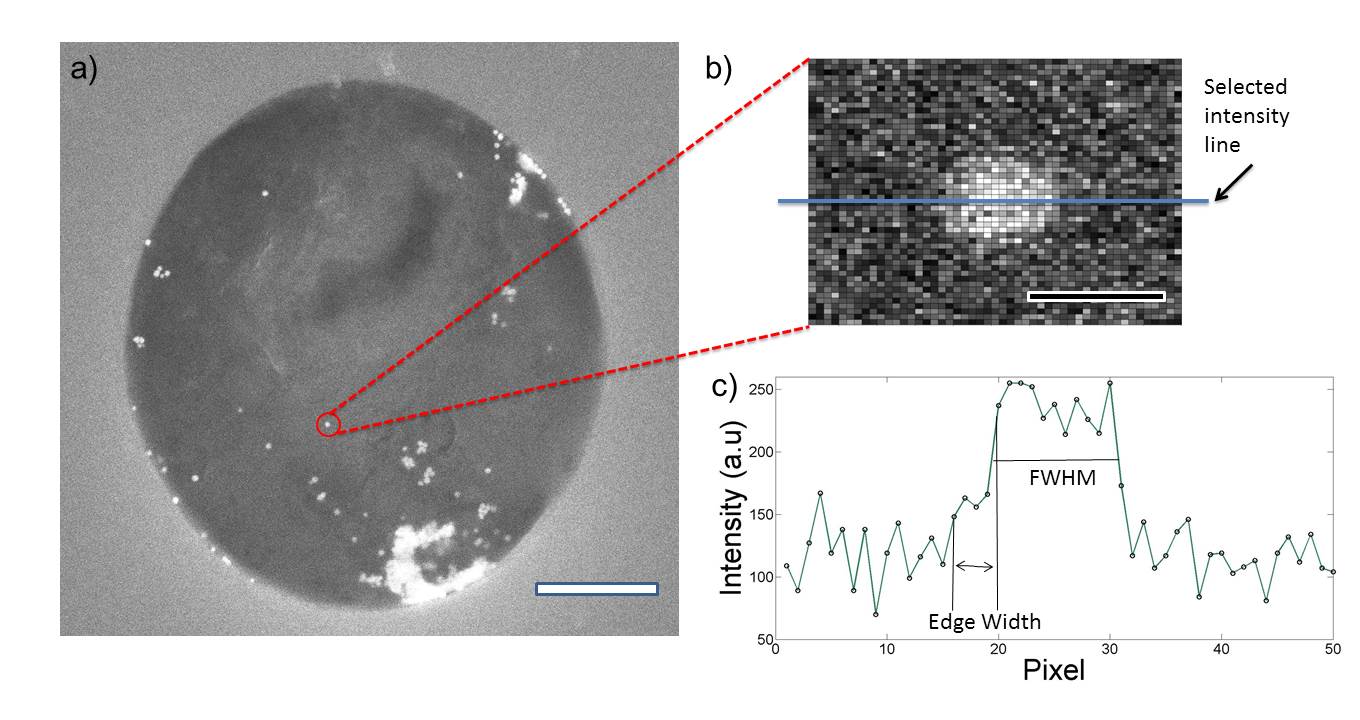}
    \caption{\small{a) A secondary electron image of 20\,nm Au particle non-specifically absorbed to the graphene window.  b) A close-up image of a bead selected as an example. c) Intensity line profile. The edge width is determined from the 20-80\% rise in intensity at the profile edge.  A resolution of 5 $\pm$ 3 nm with a CNR of 7 $\pm$ 1 is obtained from averaging results over five beads. The white and black scale bars in a) and b) corespond to 500\,nm and 25\,nm respectively.} }
    
    \label{beadresolution}
\end{figure}

To determine the resolution of the Au-NP's under the graphene wet cell we imaged 20\,nm beads bound to the membrane. The intensity line profile of each Au-NP (see Figure ~\ref{beadresolution}) was extracted and the resolution was determined from the edge-width over which the Au-NP's intensity rose from 20\% to 80\% of its maximum height\cite{de2009electron}. Averaging over five beads, we find the resolution of the Au-NP's in our wet-cell to be $5\pm3$\,nm (error is standard deviation on mean over beads measured).  

The contrast to noise ratio (CNR) is defined as :

\begin{equation}
CNR = \frac{S}{\sigma_{n}}
\end{equation}

where S is the peak signal and $\sigma_{n}$ is the standard deviation of the background noise \cite{welvaert2013definition}. We obtained a value of 7 $\pm$ 1.  These results confirm that graphene leads to improved resolution:  our resolution is higher than the 20\,nm reported in the 50 nm silicon nitride membrane and comparable to the resolution ($\sim$ 5\,nm) obtained under much higher electron imaging conditions such as at 200 Kev in a TEM \cite{de2009electron}.  The high contrast to noise (CNR) ratio also makes it possible for us to observe and record movements in the liquid environment.

\section{Conclusion}

     In conclusion, we have demonstrated that single layer CVD grown graphene is very promising for SEM based wet cell imaging, enabling dynamic imaging of Au-NP undergoing brownian motion in aqueous solution and EDX measurements in liquid.  In particular, our wet cell can be used in a conventional SEM without the need for instrument modification.  In the future, opposed to previous graphene sandwitch studies, our wet-cell can be in principle adapted for nanofluidic experiments with nanochannels etched in place of the fluid reservoirs.   Such systems might enable nanoconfinement based single molecule manipulation combined with SEM imaging, giving rise to new types of single-molecule analytical devices based on electronic as opposed to optical imaging.   Our EDX results are particularly significant in this context:  one can envision future experiments that use biomarkers with differential chemical composition to tag a range of DNA modifications enabling more efficient multiplexing.

\begin{acknowledgement}

The authors acknowledge financial support from Fonds de recherche Nature et technologies (FQRNT). The authors thank the staff at the Facility for Electron Microscopy Research (FEMR) at McGill University for the use of their electron microscopy instrument and technical support. 
\end{acknowledgement}


\bibliography{refgraphenecell}

%% graphical table of contents.
%%%%%%%%%%%%%%%%%%%%%%%%%%%%%%%%%%%%%%%%%%%%%%%%%%%%%%%%%%%%%%%%%%%%%

\end{document}